\documentclass[onecolumn,showpacs,floatfix,11pt,nofootinbib]{revtex4}
\usepackage{graphicx}

\RequirePackage{amssymb}

\begin{document}

\title{Thermodynamical Analysis of a Black Hole with a Global Monopole Within a Class of a f(R) Gravity}

\author{F. B. Lustosa$^{1}$}
\email{chicolustosa@if.uff.br}
\author{M. E. X. Guimar\~aes$^{1}$}
\email{emilia@if.uff.br}
\author{Cristine N. Ferreira$^{2}$}
\email{crisnfer@pq.cnpq.br}
\author{J. L. Neto$^{3}$}
\email{jlneto@if.ufrj.br}

\affiliation{1. Instituto de F\'{\i}sica, Universidade Federal
Fluminense, Av. Gal. Milton Tavares de Souza, s/n - Campus da
Praia Vermelha 24210-346 Niter\'oi, RJ, Brazil}

\affiliation{2. N\'ucleo de Estudos em F\'{\i}sica,  Instituto Federal de Educa\c{c}\~{a}o,
Ci\^encia e Tecnologia Fluminense,
Rua Dr. Siqueira 273, Campos dos
Goytacazes, 28030-130
RJ, Brazil}

\affiliation{3. Instituto de F\'{\i}sica, Universidade Federal do Rio de Janeiro, Caixa Postal 68528 Rio de Janeiro 21941-972, RJ, Brazil}

\pacs{?}

\begin{abstract}
We analyze the thermodynamics of a black hole in a region that contains  a global monopole in the framework of a particular class of a $f(R)$ gravity. Specifically, we study  the case in which  $\frac{df(R)}{dR} = F(R) $ is  a power law function of the radial coordinate of the monopole spacetime, i.e., $F(r) = 1 + \psi_n r^n$, where $\psi_n$ is the fine-tuned parameter corresponding to the $f(R)$ gravity. We  obtain explicit expressions for the local thermodynamic quantities of the black hole as a function of the event horizon,  the parameter describing the monopole and  the measurable corrections  due to the $f(R)$ theory modifications of the General Relativity.  
We  also discussed the implications of the particular case of $n=2$, where the parameter $\psi_2$ can be related to a positive cosmological constant, that in monopole presence is characterized by a non-trivial topology observed as a deficit solid angle. \end{abstract}
\maketitle

\newpage

\section{Introduction}

There are still some important  issues for which current standard cosmological theories in Physics have not yet found complete answers.  General Relativity (GR) has not been able to clarify questions such as  the non-renormalization of the gravity theory, the singularity problems in black hole physics and the physics of the early Universe,  leading us to the necessity of finding alternative approaches. In this context, one of the most intriguing facts is the accelerating Universe. Despite of the existence of some alternative approaches  to explain this one without adding dark matter or dark energy \cite{Buchdahl:1983zz,Morais:2015ooa,Takahashi:2015ifa}, the f(R) gravities \cite{Nojiri:2010wj,Sotiriou:2008rp} have received much attention as one of the strongest candidates to explain the current accelerating universe \cite{Perlmutter:1998np}. The f (R)  gravity is constructed by replacing the Ricci scalar in the Einstein-Hilbert action by an arbitrary function of the Ricci scalar f (R). Such theory is well known to lead to an extra scalar degree of freedom and it has been shown, for example in \cite{Hu:2007nk,Appleby:2007vb,Starobinsky:2007hu}, that observationally acceptable models can be constructed. In some new contribute the f(R) gravity can be dependence of high orders of curvature scalar \cite{Ohta:2015efa}.

In 1972, Bekenstein published his first article demonstrating  the relation between thermodynamic quantities and gravitational properties of a black hole. His works \cite{Bekenstein:1972tm}-\cite{Bekenstein:1974ax} were followed by the model of particle creation around BH proposed by Hawking \cite{Hawking:1974sw}. Since then, the thermodynamical  behavior of quasi-classical systems has been widely explored in others gravitational frameworks. Most recent contributions were made in the study of the thermodynamics of the black hole in  modified Schwarszchild \cite{Cai:2009ua}, Born-Infeld-anti-deSitter \cite{Myung:2008eb},  Horava-Lifshitz \cite{Biswas:2010zzb} and  $f(R)$ theory \cite{delaCruzDombriz:2009et}.

It has been shown that topological defects such as  cosmic strings, monopoles and domain walls
could be formed as a result of spontaneously broken symmetry in a vacuum phase transition of the early universe \cite{Kibble:1976sj}-\cite{Vilenkin:1984ib}. In the context of the GR, Barriola and Vilenkin \cite{Barriola:1989hx} studied the gravitational effects of a global monopole as a spherically symmetric topological defect. In the context of f(R) theories, global monopoles were studied recently by \cite{Carames:2011uu}-\cite{Carames:2011xi}, and references therein. In this solution, there appears a term which corresponds to a black hole solution.
It is natural, then, to imagine that it exists a region with a global monopole that have been swallowed by a black-hole.  The other interesting work where the monopole is the important ingredient is in the boson star context.  In this case when these compact objects are present with a global monopole in non-minimal coupling  a balck hole can be mimicker. This approach is important to revel the non linear gravitational effects and gravitational backreation \cite{Marunovic:2013eka}.

For the other side, the analysis of the thermodynamical properties of that case in the context of GR was made in \cite{Yu:1994fy} and recently  in the context of a f(R) theory in \cite{Man:2013sf}, where  the authors used the weak field approximation solution  presented in \cite{Carames:2011uu} with an specific {\it Ansatz} for $ f'(R) = F(R) = \frac{df(R)}{dR} $ as a linear function of the radial coordinate.

In this work we consider the thermodynamic aspects of  black holes in the spacetime with a global monopole in the f(R) context.  
We study  a general case in which the  F(R) function is  a power law function of the radial coordinate. We anticipate that we obtain the explicit expressions for the local thermodynamical quantities of the black hole as a function of the event horizon, the parameter describing the monopole and the measurable corrections on the GR gravity due to the $f(R)$ theory.

The paper is organized as follows.
In the Section 2, with the motivation given by thermodynamics aspects, we revisited the global monopole solution in the f(R) theory when  the weak field approximation takes into account \cite{Carames:2011xi}. The new features of this section are the analysis about some aspects of the Hawking thermodynamics,  the black hole and event horizons analysis, the behavior of the metric components  on positive  values of $\psi_n$  in our approximation, and we analyze the dependence of the energy (GM) of the black hole as a function of the entropy. In the Section 3, we  study the thermodynamical properties of the black hole  in a $f(R)$ global monopole spacetime. Finally, in Closing  Remarks,  we   discussed the stability of our framework and analyzed with more detailed  the particular case of $n=2$, where the parameter $\psi_2$ can be related with  a positive cosmological constant, that in monopole presence contain a  deficit solid angle.

\section{Field Equations Solution  for the $f(R)$ Gravity in the Metric Formalism}

In this section we review the solution of the field equations for a f(R) theory with spherical symmetry spacetime. We have shown that it is possible to find a black hole solution in a global monopole region by using a f(R) modification of the GR gravity with a weak field approximation.  For a f(R) theory, the action associated to the matter field coupled with gravity is given by:
\begin{equation}
S = \frac{1}{2\kappa}\int{d^4x\sqrt{-g}f(R)} + S_m(g_{\mu\nu}, \psi),\label{action1}
\end{equation}
where $\kappa = 8 \pi G $, $G$ being the Newton constant, $ g$ is the determinant of the metric $g_{\mu \nu}$ with $\mu , \nu = 0, 1, 2, 3 $, $R$ is the scalar curvature,  $S_m $ is the action associated with the matter fields and $f(R)$ is an analytic function of the Ricci scalar. In this theory the Ricci scalar in Einstein -Hilbert action is replaced by arbitrary function of Ricci scalar. The monopole which introduces  an angular deficit in the spacetime metric, gives us some interesting effects that we will discuss in next sections. Here we assume that the Christoffel symbol is a function of the metric. The general form of the time independent metric with spherical symmetry in (3+1) dimensions  is giving by
\begin{equation}
ds^2= B(r) dt^2 - A(r)dr^2 - r^2 (d\theta^2 + \sin^2\theta d \phi^2)   \label{metric1}
\end{equation}
The $g_{\mu \nu}$ metric tensor equations of motion are:
\begin{equation}
R_{\mu \nu} F(R) - {1\over 2}f(R)g_{\mu \nu} - ( \nabla_{\mu}\nabla_{\nu}-g_{\mu \nu}\Box)F(R) = \kappa T_{\mu \nu}\label{freq}
\end{equation}
with $ F(R) = {df(R)\over dR} $, and $ \Box$ being the usual notation for covariant D' Alembert operator $ \Box \equiv \nabla_{\mu} \nabla^{\mu}$.
The only term associated to the matter action is related to the global monopole described by the Lagrangean density
\begin{equation}
\mathfrak{L} = \frac{1}{2}(\partial_{\mu}\phi^a)(\partial^{\mu}\phi^a ) - \frac{1}{4}\lambda(\phi^a \phi^a - \eta^2)^2,
\end{equation}
where $\lambda$ and $\eta $ are the monopole field parameters and the triplet field that will result in a monopole configuration can be described by $ \phi^a = \eta \frac{h(r)}{r^2}x^a $, with $ a = 1, 2, 3 $ and $x^a x^a = r^2 $.
     The function h(r) is dimentionless and is constrained by the conditions $h(0) = 0$ and $h(r>\eta) \approx 1$ \cite{Barriola:1989hx}. The energy-momentum tensor associated to with that field configuration is
\begin{eqnarray}
T_{t}^{t} &=&  T^r_r \approx \frac{\eta^2}{r^2} \nonumber\\
T^\theta _\theta & = & T^\phi_\phi =0
 \label{energmoment}
\end{eqnarray}
Now we can analyze the gravitational effects of the global monopole in $f(R)$ formalism, obtaining the  trace of the field equations (\ref{freq}) in the presence of the monopole\cite{Carames:2011uu,Multamaki:2006zb} 

\begin{equation}
f(R) = {1 \over 2} R F(R) + {3\over 2} \Box F(R) - {\kappa \over 2} T\label{frcontract}
\end{equation}
where $F(R) = {df(R) \over dR}$. Substituting (\ref{frcontract}) in (\ref{freq}) we have
\begin{equation}
F(R) R_{\mu \nu} - \nabla_{\mu} \nabla_{\nu} F(R) - T_{\mu \nu} = g_{\mu \nu} \, C \label{solutionequation}
\end{equation}
where the quantity $C= {1\over 4}(F(R) \,R - \Box F(R) - \kappa T)$ is a scalar quantity that is the same to all components.

In a spherical symmetric spacetime with the metric given by (\ref{metric1})  and with the energy-momentum tensor given by (\ref{energmoment}) we find through (\ref{solutionequation}) the equations as
\begin{eqnarray}
2r F'' - \beta (r F' + 2 F) =0,  \label{frn}\\
4B(A-1) + r\Big( 2 {F' \over F} - \beta \Big) (r B' - 2 B) + 2 r^2 B'' - {4 \alpha A B  \over F} =0\label{brn}
\end{eqnarray}
where $\alpha$ and $\beta$ are:
\begin{eqnarray}
\alpha &=& 2 \kappa \eta^2, \, \,  \mbox{with}  \, \, \kappa = 8\pi G \\
\beta &=& {(AB)' \over AB} \label{beta} \,\, .
\end{eqnarray}

 We make the  assumption with time independent  solutions, i.e. $B = B(r)$ and  $A = A(r)$ give us the metric as  (\ref{metric1}) for these assumptions  we consider $F(R) = F(r)$ and $F'$ and $F''$ are the first and the second derivatives with respect to $r$, respectively.
The solution to these equations is the exact description of the global monopole in $f(R)$ theories. 
However this prescription has an analytical solution only in the weak field approximation.  In this approximation the field equations are
\begin{eqnarray}
F(R) = F(r) = 1 + \psi(r) \nonumber \\
B(r) = 1 + b(r),  \, \, \, A(r) = 1 + a(r)
\end{eqnarray}
with $\vert b(r)\vert $, $\vert a(r)\vert $ and $\vert \psi(r)\vert$ much smaller than one.  These redefinitions can be used in equations (\ref{frn}) and (\ref{brn}), and using
\begin{eqnarray}
{F' \over F} \sim \psi',& \, \, \, &  {F'' \over F} \sim \psi''\\
{A' \over A} \sim a', &\, \, \, & {B' \over B} \sim b' \label{ab}
\end{eqnarray}

The field equations (\ref{frn}) and (\ref{brn}) can then be written in a linear form as:
\begin{equation}
\beta \sim r \psi '' (r)\label{beta2},
\end{equation}
\begin{equation}
2a(r) - 2r\psi ''(r) + r \beta + r b '' (r) - 2 \kappa \eta^2 =0 ,\label{abequation}
\end{equation}
where  $\beta $ has been defined in (\ref{beta}). This is the solution for our equations in the weak field approximation for a metric with a spherical symmetry in a f(R) theory with $\frac{df(R)}{dR} = 1 + \psi(r)$.

Following ref. \cite{Carames:2011xi},   we assume that  the function
$\psi(r)$  is a power law-like function of the radio coordinate, namely $\psi(r) = \psi_{(n)} r^n$, where $\psi_n$ is a constant parameter in r.  Note, however that  in our work it has  a  n dependence. 
It is convenient to impose $\psi(r)$ as being regular
at the origin which implies that $\psi(r) = 0$.
This assumption automatically rules out all the
negative powers of n.  With this ansatz we can find solutions for the equations above. Noting that equation (\ref{beta})  with (\ref{ab}) can be rewritten as:
\begin{equation}
a' (r) + b' (r) = n(n-1)\psi_n r^{(n-1)}
\end{equation}
we can solve (\ref{abequation}) for b(r), obtaining:
\begin{equation}
b(r) = \frac{c_1}{r} - c_2\, r^2 - \psi_n r^n - \kappa \eta^2. \label{linb}
\end{equation}

We define the integration constants  $c_1 = -2GM$ and  for convenience we take $c_2 =0$ this term correspond the cosmological constant term. We will return to discuss this question in the remarks where we analyzed the $n=2$ case where the cosmological constant is naturally came of  our $f(R)$ framework.
Equations (\ref{beta}) and (\ref{beta2}) give the relation:
\begin{equation}
A(r)B(r) = a_0 e^{(n-1)\psi_n r^n}.\label{AB}
\end{equation}
Defining the integration constant $a_0 = 1$ we then have the full form of the metric as
\begin{equation}
ds^2 = (1 - \frac{2GM}{r} - \psi_n r^n - \kappa \eta^2)dt^2 - e^{(n-1)\psi_n r^n}(1 - \frac{2GM}{r} - \psi_n r^n - \kappa \eta^2)^{-1}dr^2 - r^2d\Omega^2, \label{metricfinal}
\end{equation}
In this work we consider the case where $\psi_n>0 $. The above results satisfies the stability conditions:
 i) no ghosts given by ${\partial f(R) \over \partial R} > 0 \, $; ii) no tachyons if $ {\partial^2 f(R) \over \partial R^2} > 0$ \cite{Appleby:2007vb} and  we have:
\begin{equation}
B(r) = 1 - \frac{2GM}{r} - \psi_n r^n - \kappa \eta^2 \label{B}
\end{equation}
\begin{figure}[htb]
\includegraphics[width=8cm, height=7cm]{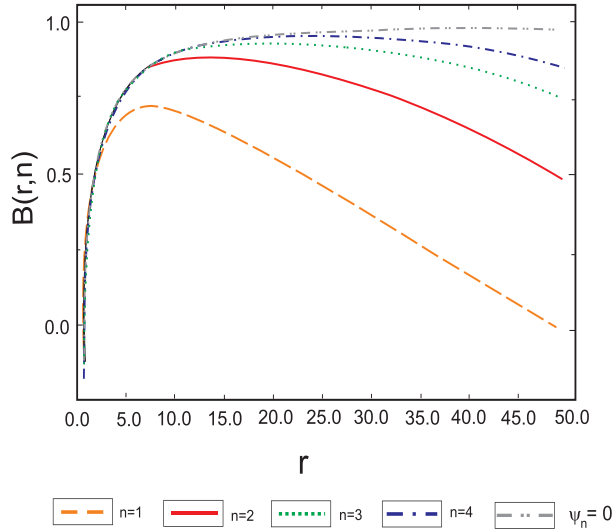}
\caption{Graph for B(r, n) for  each n- orders going from 1 to 4 and for the Schwarzschild Black Hole (SBH), where $GM=1$, $\kappa \eta^2 = 10^{-5}$ and $\psi_n =0. 2 \cdot 10^{-2n+1} $}
 \label{Figure1}
 \end{figure}
     Considering the mass term as cosmologically relevant is equivalent of considering the scenario with a black hole in the spacetime of  a global monopole  \cite{Barriola:1989hx}. To confirm that our solution is in fact a black hole type solution for all degrees $n$ of the $\psi(r) $ function, we have to find if all of them have an event horizon.
     To do this, we have to keep in mind the approximations that we used to solve the field equations, namely $h \approx 1 $, which we used to define the energy-momentum tensor outside the core of the global monopole, and $\psi(r) << 1 $. These approximations require that we restrict our analysis to a region defined by:
\begin{equation}
\delta < r < \frac{1}{\vert \psi_n r^n \vert} \label{rlimits}
\end{equation}
where $\delta \approx (\lambda\eta^{1/2})$ is the order of magnitude of the monopole's core. Assuming that  $ e^{(n-1) \psi_nr^n} \sim 1+(n-1) \psi_n r^n$ in (\ref{metricfinal})
we can write the scalar curvature in the general form as 
\begin{eqnarray}
R &=&   -3n(n+1)\psi_n r^{n-2} -2\kappa\eta^2 r^{-2}  \label{RRG1}
\end{eqnarray}

Analytically we can prove that the metric has an event horizon for the solution $n + 1$ degree equation:
\begin{equation}
1 - \frac{2GM}{r_{H}} - \psi_n r^n_{H} - \kappa \eta^2 = 0.\label{eqrh-mass}
\end{equation}
In Figure \ref{Figure1}  we plot the function $B(r,n)$  for all $n's$ where we included the Schwarzschild case. We can see that
for n going from 1 to 4 the horizon appears in the same region. 

We have written the parameter $\psi_n $  as an exponential in $n$.  This is essential to respect the condition (\ref{rlimits}) without narrowing  the region that we are going to analyze. Considering only the region where $\psi (r) << 1 $ implies that the other roots of equation, (\ref{eqrh-mass}) are out of the region where our solution is valid.
 
\begin{figure}[htb]
\includegraphics[width=8cm, height=7cm]{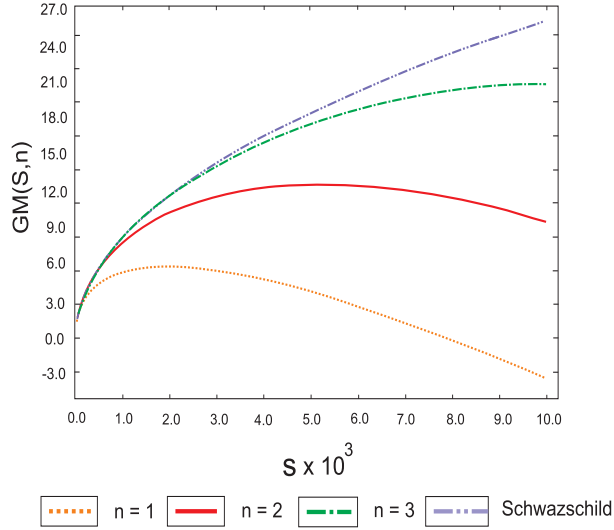}
\caption{Graph for $GM(S, n)$ for n going from 1 to 3 and for the Schwarzschild Black Hole,  $\kappa \eta^2 = 10^{-5}$ and $\psi_n =0. 2 \cdot 10^{-2n+1} $. }
\label{Figure2}
 \end{figure}
From  (\ref{eqrh-mass}),  we get  the dependence of the energy (GM) on the horizon $r_H$ 
 \begin{equation}
 GM = {1 \over 2} \Big[ (1-\kappa \eta^2)r_H - \psi_n r^{n+1}_H  \Big ]\label{mass2}
 \end{equation}
 In  the Figure \ref{Figure2}  we plotted the energy with respect with entropy of the black hole.

The Hawking  temperature can be  easily obtained from the GM expression  which correspond to the internal energy of the black hole. The Hawking temperature is, therefore,   the  derivative of the energy with respect with entropy of the black hole $S = \pi r_H^2$,
  \begin{eqnarray}
T_H(\eta, \psi_n) =  \frac{1}{4\pi}[\frac{1 - 8\pi G\eta^2}{r_H} - (n+1)\psi_n r_H^{n-1}], \label{Templocal}
\end{eqnarray}
which, for $n = 1$, recovers the result obtained in \cite{Man:2013sf}, and, for $\eta = 0$, $\psi(r) = 0$, recovers the Schwarzschild result.

\section{The thermodynamics of the black hole in a $f(R)$ Global Monopole for Generic $n$.}

\begin{figure}[h]
    \centering
    \includegraphics[width=8cm, height=7cm]{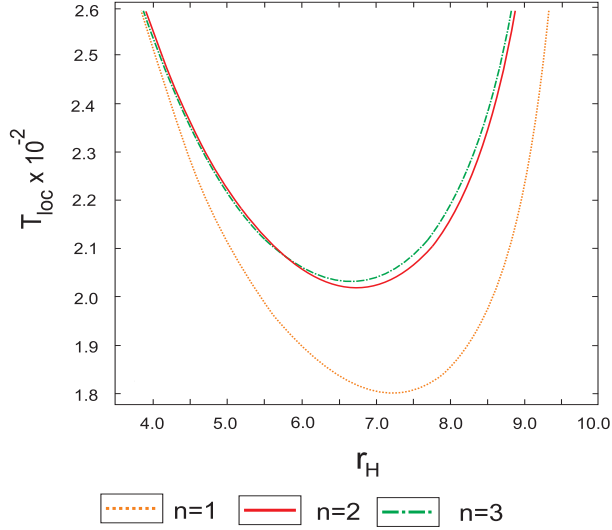}
    \caption{Graph for $T_{loc}(r_H)$ for $n=1, 2,  3$ and for the Schwarzschild black hole.}
    \label{Figure3}
  \end{figure}

In this section we will compute the thermodynamical quantities of a black hole for any value of $n$.
The Hawking temperature  becomes rapidly negative, as we show in (\ref{Templocal}) and, therefore, we should turn to a local analysis. We use the prescription given in \cite{Tolman:1930zza} to obtain a local temperature we have:
\begin{equation}
T_{loc} = \frac{1}{4\pi}\Big[\frac{1 - 8\pi G\eta^2}{r_H} - (n+1)\psi_n r_H^{n-1}\Big]\sqrt{\frac{r}{\psi_n r_H^{n+1} - r_H(1 - 8\pi G \eta^2) + r(1 - 8\pi G \eta^2) - \psi_n r^{n+1}}}\, .\label{Tloc}
\end{equation} 
     The local temperature for $n=1$ to $3$ and  the Schwarzschild black hole is plotted in figure \ref{Figure3} as a function of the event horizon where we fixed the position at $r = 10$. The graph is zoomed  to show the vary slight difference between the minimum values of the temperature for different values of the degree $n$. The only notable difference is between n=1 and the others values of n.

This figure 3 also  shows that  for all values of n there are two black holes one where the temperature diminished with the horizon $r_H$ and the other when the temperature grows with the horizon. The solution that give us the temperature decreasing with the horizon is stable and is the so called big black hole. The other, the small black hole is unstable. This result will become more clear when we  present the heat capacity of the  black hole.

\begin{figure}[htb]
\includegraphics[width=9cm, height=7cm]{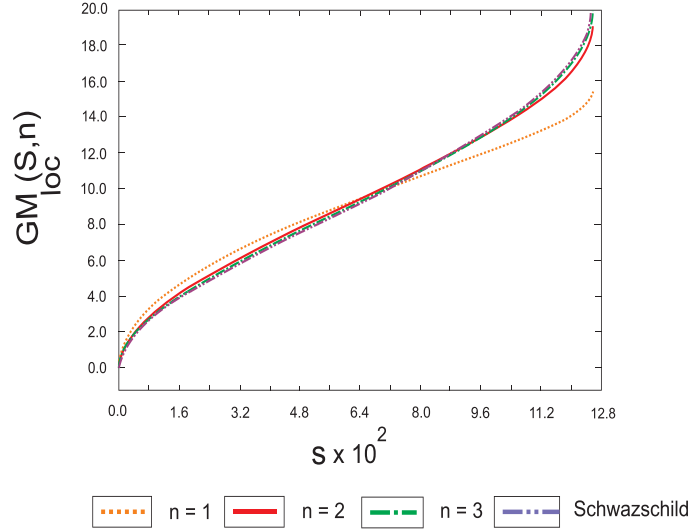}
\caption{Graph for $GM(S, n)_{loc}$ for n going from 1 to 3  r= 20,  $\kappa \eta^2 = 10^{-5}$ and $\psi_n =0. 2 \cdot 10^{-2n+1} $. }
\label{Figure4}
 \end{figure}

From the first law of thermodynamics  $ dGM_{loc} = T_{loc} dS$, the thermodynamical local energy $GM_{loc}$ can be derived. With  the integration constants were chosen conveniently, we get
\begin{eqnarray}
GM_{loc} &=&  r\sqrt{(1 - 8\pi G \eta^2 - \psi_n r^n)} - \sqrt{r}\sqrt{r(1 - 8\pi G \eta^2 - \psi_n r^n) - r_H(1 - 8\pi G \eta^2 - \psi_n r_H^n)}\ ,\label{Eloc}
\end{eqnarray}
 In  Figure \ref{Figure4} we plot the Energy in function of entropy.  It can be seem  that  in the local framework the energy and temperature T  are positive.

We will  now look after the local heat capacity, which can be calculated from the energy. The expression is obtained as follows:
\begin{equation}
C_{loc} = (\frac{d GM_{loc}}{d T_{loc}})_r = (\frac{d GM_{loc}}{d r_H} \cdot (\frac{d T_{loc}}{d r_H})^{-1})_r
\end{equation}
The subscript is there to show that we are calculating this quantity in a fixed position. Inputting equations (\ref{Tloc}) and (\ref{Eloc}), we obtain the complicated expression:
\begin{equation}
C_{loc} = \frac{H(r_H, r, \eta, n)}{W(r_H, r, \eta, n)},
\end{equation}
where
\begin{eqnarray}
H(r _H, r, \eta, n) =  2\pi r_H^2(1 - 8 \pi G \eta^2 - (n+1)\psi_n r_H^n) \nonumber  \\
 \times [r(1 - 8 \pi G \eta^2 - \psi_n r^n) - r_H(1 - 8 \pi G \eta^2 - \psi_n r_H^n)], \\
W(r, \eta, n) = \frac{r_H}{2}(1 - 8 \pi G \eta^2 - (n+1)\psi_n r_H^n)^2  \nonumber \\
 - (1 - 8 \pi G \eta^2 + (n^2-1)\psi_n r_H^n)(r(1 - 8 \pi G \eta^2 - \psi_n r^n) - r_H(1 - 8 \pi G\eta^2 - \psi_n r_H^n).
\end{eqnarray}

\begin{figure}[htb]
    \centering
    \includegraphics[width=9cm, height=8cm]{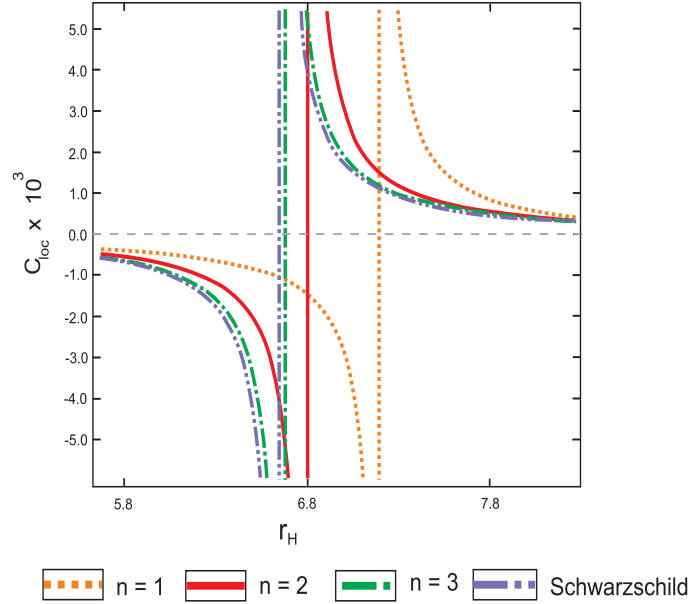}
    \caption{Graph for $C_{loc}(r_H,n)$ for the Schwarzschild case and for the $n = 1$, $n = 2$ and $n = 3$ and the Schwarzschild black hole cases, for fixed r = 10.}
    \label{Figure5}
  \end{figure}

\begin{figure}[h]
    \centering
    \includegraphics[width=9cm, height=7cm]{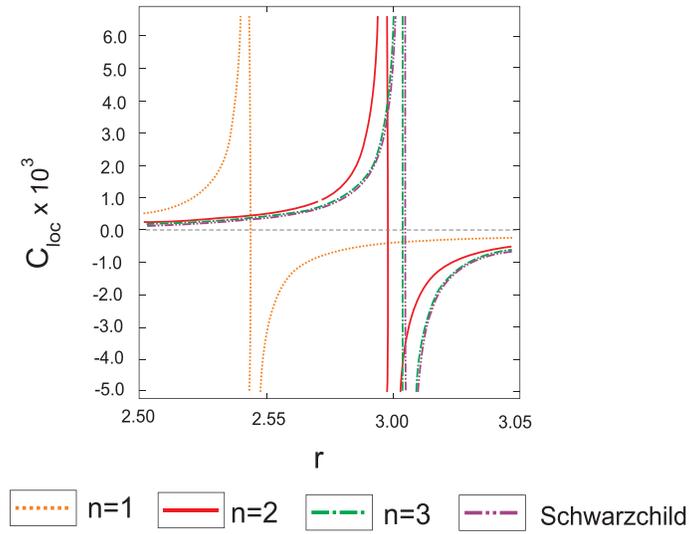}
    \caption{Local heat capacity as a function of position for n=1, 2, 3 and the Schwarzschild black hole. }
    \label{Figure6}
  \end{figure}

\begin{figure}[h]
    \centering
    \includegraphics[width=8cm, height=7cm]{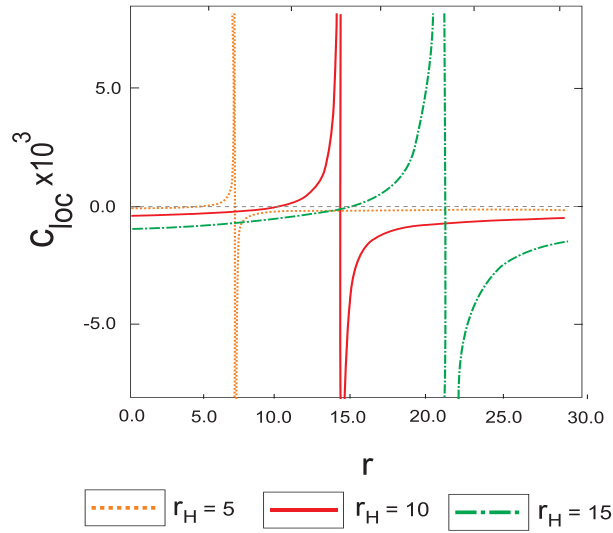}
    \caption{Local heat capacity for n= 2 as a function of position with $r_H = 5, 10, 15$.}
    \label{Figure7}
  \end{figure}

  In figure \ref{Figure5}, we see that the local heat capacity plotted for the classical case and the $n = 1$ to  $3$ and  the local form for the Schwarzschild black hole. The curves that are slightly distinguishable are  $n =1$ and $n= 2$, the other ones are superposed near to the Schwarzschild black hole.
The Figure \ref{Figure5}  also shows that the heat capacity refers to a specific point near the horizon. The transition from negative to positive values happens as the horizon is closer to the chosen position $r$. We can see this feature more clearly by analyzing the heat capacity as a function of the position, with a fixed horizon, as is done in Figures \ref{Figure6}  and \ref{Figure7} .
     The expected negative heat capacity will be observed as $r$ grows and the transition to positive values happens even for bigger black hole. This transition is yet to be explained and it occurs even for the classical Schwarzschild case in the local analysis, but not in the most usual global calculation where it remains negative for every value of $r_H$.

\newpage

\section{Closing Remarks}

 In this work, we have analyzed the solution of a black hole in the spacetime of a global monopole in a class of $f(R)$ gravity and the thermodynamics properties that came from this model in weak field approximation.  In our approach, we have considered  the scalar function $\frac{df(R)}{dR} = F(R)  =1+ \psi_n \, r^n $ assuming the   {\it ansatz} that the curvature scalar $R$ is a function of the radial coordinate only.  
 
It is important here to make closing remarks about the stability of the f(R) function that it can be analyzed for each n order in our power law expansion. The stability discussion in the case $n=1 $ with the BH can be made as presented in \cite{Carames:2011uu} and in 
   \cite{Man:2013sf} where the authors study its thermodynamical properties.  
 The new feature is the n=2 case where  the parameter $\psi_2$ can be related with  a positive cosmological constant, that in monopole presence is characterized by a non-trivial topology observed as a solid deficit angle.  We consider in (\ref{RRG1})  $ n=2$ that give us  $
R= -18\psi_2  -2\kappa\eta^2 r^{-2} $.
If we now compare this curvature scalar to General Relativity  we see that the parameter $\psi_2$ can be identified with the cosmological constant  as $\psi_2 = {2 \over 9} \Lambda$.

Notice that the resulting $f(R)$ gravity is not unique due to the presence of the integration constant given by the solution of the  equation ${df(R) \over dR} = F(R) = 1-  {\alpha \psi_2 \over R +18\psi_2}$  by integration we can write the expression of
$f(R)$ as
\begin{equation}
f(R) = R - \alpha \psi_2  \ln\Big(R +18\psi_2 \Big) + L \,\, , \label{frsolution}\end{equation}
where L is the integration constant.  Therefore to specify L, we impose that $f(0) = 0$. Hence we find $L = \alpha \psi_2  \ln\Big(18\psi_2 \Big)$. When $R\rightarrow \infty$ $f(R)/R \rightarrow 1$.  Writing $f(R) = R + \epsilon(R)$, this implies that  $\epsilon/R \rightarrow 0$ as $R \rightarrow \infty $. This in turn implies that $f'(R) \rightarrow 1$ as $R\rightarrow \infty $ from which we can deduce that $\epsilon(R) \rightarrow $  constant as $R\rightarrow \infty $.  This f(R) function also satisfy in n=2 case the stability function given by: i) no ghosts given by ${\partial f(R) \over \partial R} > 0 \, $; ii) no tachyons if $ {\partial^2 f(R) \over \partial R^2} > 0$ in the case where $\psi_2 >0$. The same procedure applies to the other values of $n$  giving us an $n$- order polynomial equation implying that the function $F(r)$ has
$n$ roots but only one of these roots satisfies the stability conditions. We can see that the $f(R)$ modification (\ref{frsolution}) has an explicit dependency on the monopole parameter $\alpha = 2\kappa \eta^2  $ and on the $f(R)$ parameter $\psi_n$ and we can observe that when $\eta = 0 $ the f(R) becomes General Relativity with positive cosmological constant with: the temperature, energy and heat capacity recover the Schwarzchild Black Hole with cosmological constant $ \Lambda= {9\over 2} \psi_2$. In this form we show that the f(R) solution is generated by the
monopole potential. 

After we have done the stability discussion in this framework considering the n=2 case  that mimics the positive cosmological constant that arrived  naturally of the f(R) theory without consider additional parameters. 

We analyze the thermodynamic effect near of the event horizon.  We observe with the plot shown in figure 3 that for $n>2$ there are no relevant modifications in the graph of the local temperature with the horizon  $r_H$. This plot showed two regions one where the temperature diminished with the horizon $r_H$ that gives us the stable black hole that corresponds to the negative heat capacity that is compatible with the black hole results. This behavior occurs for all n's remembering that the relevant one is when n=2, that corresponds to the case of the positive cosmological constant with the solid defict angle as we analyzed. 
The local analysis of the thermodynamic quantities of the BH indicates that some kind of observable phase transition takes place in regions near the event horizon, and this might indicate some interesting thermodynamical  process without matter.  The other region presents the temperature growth with the horizon, this one corresponds to the unstable black hole and corresponds to the positive heat capacity.

 In this work we have not analyzed the case in which the  n=2  for $\psi_2<0$. In  future works, we will analyze this case that might be  important for the  AdS/CFT context for n=2 case. These  aspects are being analyzed and will be the subject for another work. It is possible to solve the stability problem of a negative $ \psi_2$ and to use the same thermodynamic procedure used here. The importance, in the AdS context,  is the fact that this black hole presents an angular deficit, and it is important to study the holographic principle of this object. We showed in \cite{Bayona:2010sd} that the structure of the angular deficit in a theory in $D=(3+1)$ is preserved on the boundary $(2+1)$ and in our case, the defect are preserved in $D=(2+1)$. It is interesting when analyzing this case for the monopole configuration. In f(R) theory the stability is solved if we put in (\ref{AB}) with the integration constant $a_0 \neq1$, in this case a $r^{-2}$ term appears and the curvature scalar change give us combination between $a_0$ and $\psi_2$  \cite{Multamaki:2006zb}.  The $AdS_4/CFT_3 $ is very important in studying planar materials like graphene and topological insulators.  The introduction of the defect in this structure gives us current properties in these materials and is an important feature for our work, giving us a lot of application possibilities for our model.  The other issue that we can analyze is the behavior of the objects near these systems \cite{BallonBayona:2013gx} with both regime for $\psi$. We can study  intensively  the pair quark anti-quark and study the confining and non-confining transitions \cite{BallonBayona:2008uc}.

\vspace{1 true cm} { \bf ACKNOWLEDGEMENTS: }
The authors would like to thanks T.R. P. Caram\^es for interesting discussions. C.N.F.  would like to thank the CNPq/Brazil for a support.

\end{document}